\documentclass[aps,prl,reprint]{revtex4-2}

\usepackage[justification=justified,singlelinecheck=false]{caption}
\usepackage{float}
\usepackage{amsmath,amssymb,amsfonts}
\usepackage{graphicx}
\usepackage{overpic}
\usepackage{bm}
\usepackage[left]{lineno}
\makeatletter
\renewcommand{\paragraph}[1]{%
  \par\vspace{0.6em}%
  \noindent\textit{#1 ---}
}
\makeatother

\usepackage{color}

\usepackage{braket}
\usepackage{esint}
\usepackage{dutchcal}
\usepackage[utf8]{inputenc}

\usepackage{xcolor}

 \usepackage[unicode=true,pdfusetitle,
bookmarks=true,bookmarksnumbered=false,bookmarksopen=false,breaklinks=true,pdfborder={0 0 1},backref=false,colorlinks=true]
  {hyperref}
 \hypersetup{
 linkcolor=red,urlcolor=blue,citecolor=red,anchorcolor=blue}

\makeatletter
\@ifundefined{textcolor}{}
{%
 \definecolor{BLACK}{gray}{0}
 \definecolor{WHITE}{gray}{1}
 \definecolor{RED}{rgb}{1,0,0}
 \definecolor{GREEN}{rgb}{0,1,0}
 \definecolor{BLUE}{rgb}{0,0,1}
 \definecolor{CYAN}{cmyk}{1,0,0,0}
 \definecolor{MAGENTA}{cmyk}{0,1,0,0}
 \definecolor{YELLOW}{cmyk}{0,0,1,0}
}

\usepackage{xcolor}
\usepackage{ifthen}
\usepackage{marginnote}

\newboolean{showedits}
\setboolean{showedits}{true}  

\usepackage[normalem]{ulem}

\newif\iferbedit
\erbedittrue   
\erbeditfalse  

\iferbedit
  \newcommand{\erbedit}[1]{\textcolor{blue}{#1}}
  \newcommand{\erbdel}[1]{\textcolor{red}{\sout{#1}}}
  
\else
  \newcommand{\erbedit}[1]{#1}
  \newcommand{\erbdel}[1]{}
  
\fi

\usepackage{times}
\usepackage{psfrag}
\usepackage{epsf}
\usepackage{epsfig}
\usepackage{latexsym}
\usepackage{amsfonts}
\usepackage{graphicx}
\usepackage{subcaption}
\usepackage{color}
\usepackage{epsf}
\usepackage{latexsym}
 \usepackage{ifpdf}

\makeatother

\begin{document}


\title{Phase-Reference Control of Steady-State Entanglement in Open Quantum Systems}

\author{Areeda Ayoub} 
\affiliation{Department of Physics,
University of Houston, Houston TX, USA}

\author{Alfonso Castillo-Gonzalez} 
\affiliation{Department of Physics,
University of Houston, Houston TX, USA}

\author{Eric R. Bittner}
\email{ebittner@central.uh.edu}
\affiliation{Department of Physics,
University of Houston, Houston TX, USA}

 \date{\today}

\begin{abstract}
\erbedit{We show that steady-state entanglement in open quantum systems is controlled by the phase reference of a phase-sensitive reservoir. Using a covariance-matrix approach for Gaussian-preserving dynamics, we demonstrate that purely local, phase-sensitive dissipation can generate entanglement when combined with coherent coupling.} The steady state exhibits a finite entangled region with an optimal squeezing strength that maximizes both the magnitude and thermal robustness of entanglement. We find that coherent coupling does not enhance entanglement monotonically, but instead regulates the conversion of local squeezing into nonlocal correlations. 
\erbedit{Importantly, the coupling dependence is controlled by the phase reference of the squeezed reservoir: phase-locked (rotating-frame) and laboratory-frame implementations yield qualitatively distinct steady states and entanglement structure.} These results establish phase-sensitive reservoir engineering as a controllable route to steady-state entanglement in continuous-variable systems.
\erbedit{Steady-state entanglement in phase-sensitive open systems depends explicitly on the reservoir phase reference and is not invariant under changes of that reference.}
\end{abstract}

\maketitle

\paragraph{Introduction.}
\erbedit{Quantum entanglement is a defining feature of quantum mechanics and a central resource for quantum communication, computation, and sensing~\cite{kimble2008quantum,nielsen2010quantum,giovannetti2004quantum}. Beyond applications, it provides a direct probe of nonclassical correlations in many-body systems~\cite{amico2008entanglement,horodecki2009quantum}. A central challenge is to generate and stabilize entanglement under realistic conditions, where environmental interactions lead to decoherence and dissipation that suppress quantum correlations~\cite{breuer2002theory,weiss2012quantum,yu2009sudden}.}

\erbedit{Reservoir engineering inverts this paradigm by using dissipation as a resource~\cite{poyatos1996quantum,kronwald2013arbitrarily}. In particular, squeezed reservoirs provide phase-sensitive fluctuations that can be transferred to a system to generate steady-state coherence and entanglement~\cite{gardiner1986inhibition,clerk2010introduction}. These ideas are experimentally accessible in optical and circuit-QED platforms, where squeezed fields can be generated, controlled, and phase-locked to a system or measurement reference~\cite{slusher1985observation,mallet2011quantum}.}

Here, we consider two linearly coupled harmonic oscillators, each interacting with its own independent squeezed reservoir. Because the dissipation is strictly local, any entanglement must arise from the interplay between coherent coupling and phase-sensitive reservoir fluctuations, \erbedit{rather than} from a shared bath. This setup isolates the minimal \erbedit{mechanism by which local, phase-sensitive dissipation is converted into nonlocal quantum correlations}, while remaining directly relevant to current experimental platforms.

A central aspect of the problem is \erbedit{that phase-sensitive reservoirs are not uniquely defined without specifying a phase reference}. If the squeezing is phase-locked to the system, the anomalous correlations are stationary in the rotating frame and the steady state admits an analytical description. If instead the squeezing is fixed relative to the laboratory frame, the correlations become \erbedit{explicitly} time dependent and must be treated numerically. These \erbedit{correspond to distinct experimentally realizable phase references and lead to inequivalent steady states with qualitatively different coupling dependence}.

We analyze the steady state through the covariance matrix of the quadrature operators $(x_1,p_1,x_2,p_2)$ and determine how entanglement depends on squeezing, temperature, and intermode coupling. \erbedit{Coherent coupling does not simply enhance entanglement, but regulates the conversion of local squeezing into nonlocal correlations.} In the phase-locked (rotating-frame) regime, the steady state is bounded and exhibits a dome-like dependence of the critical temperature on squeezing strength. In contrast, fixing the squeezing in the laboratory frame produces a different ordering of \erbedit{entanglement with respect to coupling}. These results establish that steady-state entanglement in open continuous-variable systems is controlled not only by dissipation and coupling, but also by the phase reference \erbedit{used to define the reservoir}.

\erbedit{While phase-sensitive reservoirs are known to depend on a reference phase, it is not generally expected that steady-state entanglement in open systems should depend on how that reference is defined.} \erbedit{Here we show that this dependence is physical, leading to inequivalent steady states with distinct entanglement properties.}

\paragraph{Model and formalism.}
We consider two linearly coupled harmonic oscillators, each interacting with an independent squeezed thermal reservoir. The system Hamiltonian is
\begin{equation}
H_S=\omega_1 a_1^\dagger a_1+\omega_2 a_2^\dagger a_2
+J\left(a_1^\dagger a_2+a_1 a_2^\dagger\right),
\end{equation}
where $J$ denotes the coherent intermode coupling.

The reduced dynamics are described within the Born--Markov approximation by a Gorini--Kossakowski--Sudarshan--Lindblad master equation~\cite{gorini1976completely,lindblad1976generators}. Each reservoir is taken to be in a squeezed thermal state characterized by squeezing strength $r_k$ and phase $\phi_k$, which introduces both standard dissipative processes and phase-sensitive contributions arising from anomalous correlations~\cite{gardiner1986inhibition,clerk2010introduction}.

To obtain a time-independent description, we move to a frame rotating at the oscillator frequency by defining slowly varying operators $\alpha_j(t)=a_j(t)e^{i\omega_j t}$ and apply the rotating-wave approximation (RWA)\erbedit{, which eliminates nonresonant terms oscillating at $\pm 2\omega_j$}. In this frame, rapidly oscillating terms average out, and the anomalous correlations of the squeezed reservoirs become stationary provided the squeezing phase is defined relative to the system dynamics. This corresponds physically to a phase-locked configuration, as realized in experiments where the squeezing source is referenced to the system or to a measurement local oscillator. The resulting master equation is time independent and can be written as
\begin{align}
\dot{\rho}
&=
-i[H_S,\rho]
+
\sum_{k=1}^{2}
\Big[
\gamma_k(N_k+1)\mathcal{D}[a_k]\rho
+
\gamma_k N_k \mathcal{D}[a_k^\dagger]\rho
\nonumber\\
&\qquad
-
\gamma_k M_k \mathcal{S}[a_k]\rho
-
\gamma_k M_k^\ast \mathcal{S}[a_k^\dagger]\rho
\Big],
\end{align}
where
\begin{align}
\mathcal{D}[O]\rho &= O\rho O^\dagger - \tfrac{1}{2}\{O^\dagger O,\rho\}, \\
\mathcal{S}[O]\rho &= O\rho O - \tfrac{1}{2}\{OO,\rho\}.
\end{align}

The effective bath parameters are
\begin{align}
N_k &= \bar n_k\cosh(2r_k)+\sinh^2 r_k, \\
M_k &=
-\frac{1}{2}(2\bar n_k+1)\sinh(2r_k)e^{i2\phi_k},
\end{align}
where $\bar n_k$ is the thermal occupation. 
The quantity $N_k$ represents the effective population, while $M_k$ represents phase-sensitive two-photon correlations responsible for dissipative generation of entanglement \erbedit{and the transfer of correlations between modes}.

The dynamics are Gaussian-preserving, so the steady state is fully characterized by the first and second moments of the canonical quadrature operators. An equivalent description is obtained from the quantum Langevin equations for the rotating-frame operators,
\begin{align}
\dot \alpha_1 &= -\frac{\gamma_1}{2}\alpha_1 - iJ \alpha_2
+ \sqrt{\gamma_1}\, \xi_{1,\mathrm{in}}(t), \\
\dot \alpha_2 &= -\frac{\gamma_2}{2}\alpha_2 - iJ \alpha_1
+ \sqrt{\gamma_2}\, \xi_{2,\mathrm{in}}(t),
\end{align}
where $\xi_{k,\mathrm{in}}(t)$ are input noise operators with stationary correlations determined by $N_k$ and $M_k$.

We introduce the quadrature vector 
$\hat{\mathbf{X}}=(\hat{x}_1,\hat{p}_1,\hat{x}_2,\hat{p}_2)$, 
defined from the rotating-frame operators as $\hat{x}_j=(\alpha_j+\alpha_j^\dagger)/\sqrt{2}$ and $\hat{p}_j=(\alpha_j-\alpha_j^\dagger)/(i\sqrt{2})$. The steady state is fully described by the covariance matrix~\cite{braunstein2005quantum}
\begin{equation}
V_{ij}=\frac{1}{2}\left\langle \{\Delta \hat{X}_i, \Delta \hat{X}_j\} \right\rangle,
\end{equation}
where $\Delta \hat{X}_i=\hat{X}_i-\langle \hat{X}_i\rangle$. The quadratures satisfy $[\hat{X}_i,\hat{X}_j]=i\Omega_{ij}$, with $\Omega=\bigoplus_{k=1}^2 
\begin{pmatrix}
0 & 1 \\
-1 & 0
\end{pmatrix}$.

The covariance matrix satisfies the Lyapunov equation
\begin{equation}
A V + V A^{\mathrm{T}} + D = 0,
\end{equation}
where $A$ and $D$ are determined by the coherent and dissipative dynamics.

Entanglement is quantified using the logarithmic negativity. For Gaussian states, the smallest symplectic eigenvalue $\tilde{\nu}_-$ of the partially transposed covariance matrix determines entanglement~\cite{peres1996separability,horodecki1996necessary}, with
\begin{equation}
E_N=\max\left\{0,-\log_2\left(2\tilde{\nu}_{-}\right)\right\}.
\end{equation}

In the symmetric resonant limit, the covariance matrix and symplectic spectrum can be obtained analytically. The smallest symplectic eigenvalue takes the form
\begin{equation}
\tilde{\nu}_-
=
\frac{y}{2}
\left[
\sqrt{K^2-\mu^2 S^2}
-\chi S
\right],
\end{equation}
where
\begin{align}
K &= \cosh(2r), \qquad S = \sinh(2r), \\
\chi &= \frac{2\gamma J}{\gamma^2+4J^2}, \qquad
\mu = \frac{\gamma^2}{\gamma^2+4J^2}, \\
y &= 2\bar n + 1 = \coth\!\left(\frac{\omega}{2k_B T}\right).
\end{align}

The entanglement threshold $\tilde{\nu}_-=1/2$ yields
\begin{equation}
y\,R(r,J)=1,
\end{equation}
with
\begin{equation}
R(r,J)=
\sqrt{K^2-\mu^2 S^2}
-\chi S,
\end{equation}
\erbedit{Here, the term proportional to $\mu$ renormalizes the local squeezing contribution, while the term proportional to $\chi$ enters linearly and reflects the coherent transfer of phase-sensitive correlations between modes.}
The corresponding critical temperature is then given by
\begin{equation}
T_c(r,J)=
\frac{\omega}
{2k_B\,\operatorname{arctanh}\!\left[R(r,J)\right]}.
\end{equation}

The use of the rotating-wave approximation is not merely a technical simplification, but \erbedit{corresponds directly to experimentally relevant conditions}. In typical implementations, the phase of the squeezed reservoir is defined relative to an external reference, such as the pump field or the local oscillator used for homodyne detection. When this reference is phase-locked to the system, the squeezing effectively co-rotates with the oscillator dynamics, rendering the anomalous correlations stationary in the rotating frame. Under these conditions, the RWA provides an accurate and physically meaningful description of the steady state. In contrast, if the squeezing phase is fixed in the laboratory frame without phase locking, the anomalous correlations become explicitly time dependent, leading to a periodically modulated covariance matrix that must be treated numerically\erbedit{, corresponding to a periodically driven (Floquet) Gaussian steady state}. 
\erbedit{The rotating-frame formulation captures the experimentally controlled regime in which phase-sensitive reservoir engineering produces a stationary steady state with well-defined entanglement.}


\paragraph{Results.}

We analyze the steady-state entanglement of two linearly coupled harmonic oscillators interacting with independent squeezed reservoirs. Correlations are quantified using the logarithmic negativity $E_N$, obtained from the steady-state covariance matrix of the quadrature operators $(x_1,p_1,x_2,p_2)$. \erbedit{Entanglement is present whenever the smallest partially transposed symplectic eigenvalue satisfies $\tilde{\nu}_- < 1/2$.}

\begin{figure}[t]
\centering

\includegraphics[width=\linewidth]{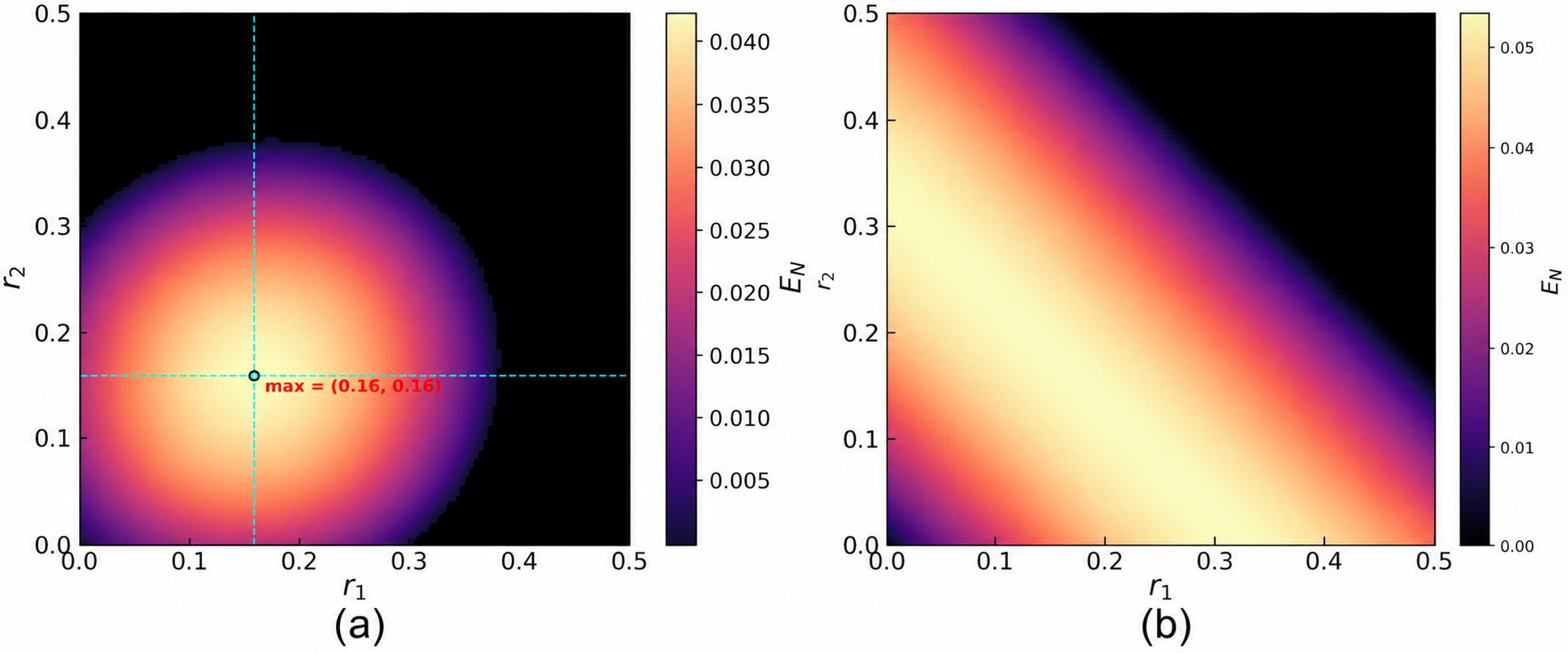}
\caption{
\erbedit{Logarithmic negativity $E_N(r_1,r_2)$ for two coupled harmonic oscillators with independent squeezed reservoirs, evaluated at $J=0.7$, $\gamma=0.5$, and $\omega=1$. (a) Laboratory-frame result. (b) Rotating-frame (phase-locked) result. The entangled region exhibits a finite optimal squeezing, reflecting competition between correlation generation and noise amplification.}
}
    \label{fig:ENmaps}
\end{figure}

Figure~\ref{fig:ENmaps}(a) shows $E_N$ as a function of the local squeezing strengths $(r_1,r_2)$. A finite entangled region emerges, separated from a separable phase by a sharp boundary. The entanglement is maximized near symmetric squeezing, $r_1 \approx r_2$, indicating that balanced reservoir engineering most effectively converts local phase-sensitive noise into shared correlations. For weak squeezing, the injected correlations are insufficient to satisfy the entanglement condition, while for strong squeezing the effective occupation increases and excess noise suppresses correlations. The resulting structure reflects a competition between correlation buildup \erbedit{($\propto \sinh 2r$)} and noise amplification \erbedit{($\propto \cosh 2r$)}, producing an optimal intermediate regime.

The phase dependence is shown in Fig.~\ref{fig:ENmaps}(b). The entanglement exhibits a pronounced stripe pattern as a function of $(\phi_1,\phi_2)$, reflecting interference between phase-sensitive dissipative channels. Maximum entanglement occurs along diagonal bands corresponding to favorable relative phase alignment, while extended regions suppress entanglement entirely. This behavior arises from interference between Liouville-space pathways associated with the anomalous terms $M_k$, which \erbedit{couple creation and annihilation processes}. 
The relative squeezing phase \erbedit{provides a direct control parameter for steady-state entanglement}.

\begin{figure*}[t]
\centering
\includegraphics[width=\linewidth]{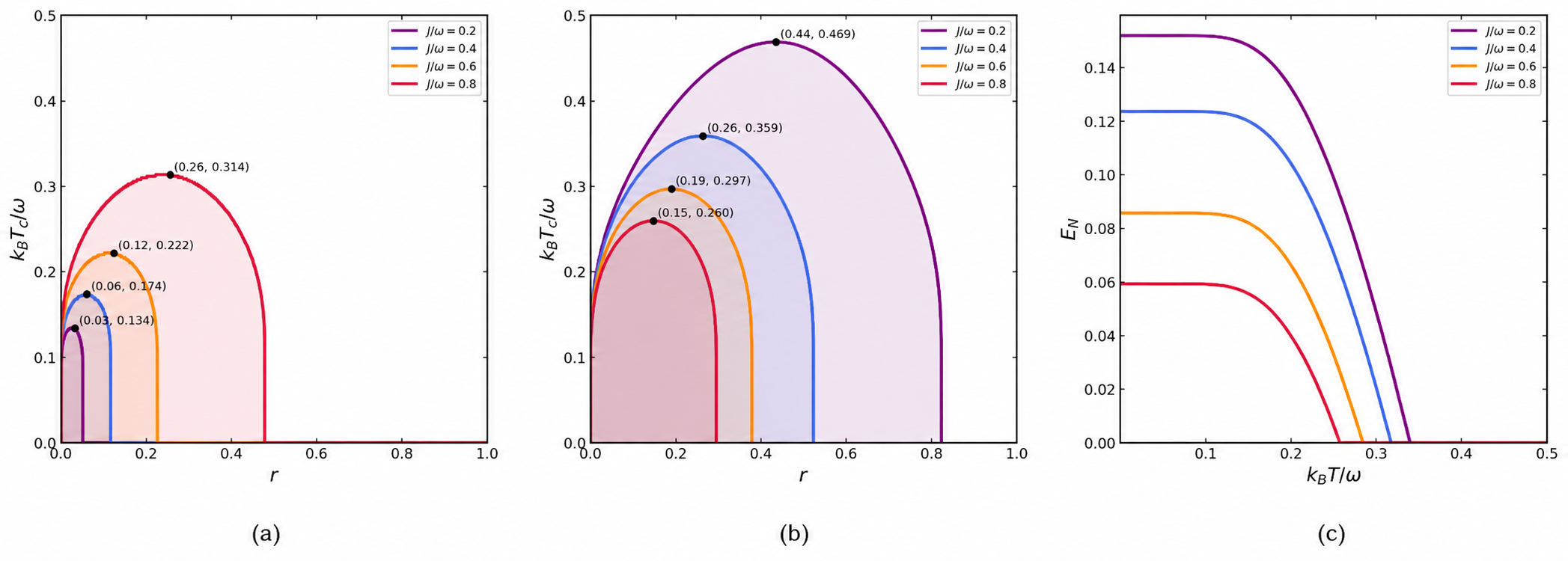}
\caption{
\erbedit{Critical temperature $T_c$ and steady-state entanglement for laboratory-frame and rotating-frame implementations of the squeezed reservoir. (a) Laboratory-frame calculation of $T_c(r)$ for several couplings $J$, where the anomalous bath correlations are fixed relative to the laboratory quadratures. (b) Rotating-frame (phase-locked) result for $T_c(r,J)$, exhibiting bounded, dome-like behavior. (c) Logarithmic negativity $E_N(T)$ at fixed squeezing, computed in the rotating frame. The ordering of $T_c$ with respect to $J$ depends on the reservoir phase reference, demonstrating that steady-state entanglement is not invariant under changes in this reference.}
}
\label{fig:frame_comparison}
\end{figure*}

Figure~\ref{fig:frame_comparison} compares two physically distinct ways of specifying the squeezed reservoir. In the fixed-frame calculation, the anomalous bath correlation is held fixed relative to the laboratory quadratures. In the rotating-frame calculation, the squeezed reservoir is phase-locked to the oscillator, so the anomalous correlations are stationary in the co-rotating frame. Both descriptions are physically meaningful, but they correspond to \erbedit{distinct experimentally realizable phase references}. 
\erbedit{This distinction leads to a qualitative reordering of $T_c$ with respect to $J$, demonstrating that steady-state entanglement depends explicitly on the reservoir phase reference.}
\paragraph{Conclusion.}
We have analyzed the steady-state entanglement of two linearly coupled harmonic oscillators interacting with independent squeezed reservoirs. Using a covariance-matrix formulation for Gaussian-preserving dynamics, we \erbedit{showed that engineered dissipation can generate and control quantum correlations in an open system}.

Steady-state entanglement emerges within a finite region of parameter space governed by the interplay of squeezing, temperature, and coherent coupling. Balanced squeezing maximizes entanglement, while the pronounced phase dependence reflects interference between phase-sensitive dissipative channels, \erbedit{identifying the relative squeezing phase as a direct control parameter}.

\erbedit{A central result is that the coupling dependence of the entanglement is not universal, but depends on the phase reference used to define the squeezed reservoir.} When the squeezing is phase-locked to the system (rotating-frame description), the steady state is bounded and exhibits a dome-like dependence of the critical temperature $T_c$ on squeezing strength. In contrast, when the squeezing is fixed in the laboratory frame, the anomalous correlations become time dependent and the ordering of $T_c$ with respect to coupling is modified, \erbedit{corresponding to a periodically driven (Floquet) Gaussian steady state}. These two cases correspond to distinct experimentally realizable phase references and lead to qualitatively different entanglement structure.

Our results show that coherent coupling does not simply enhance entanglement, but regulates the conversion of local squeezing into nonlocal correlations. While finite coupling is required to distribute phase-sensitive fluctuations between modes, stronger coupling hybridizes the system and limits the buildup of quadrature correlations.

The model is experimentally accessible in optical and circuit-QED platforms where squeezed reservoirs and tunable couplings are available~\cite{slusher1985observation,wu1986generation,mallet2011quantum,flurin2012generating,eichler2011observation}. \erbedit{More broadly, steady-state entanglement in phase-sensitive open systems depends explicitly on the reservoir phase reference and is not invariant under changes of this reference.} This provides a concrete route for tailoring quantum correlations through phase-sensitive reservoir design.



\begin{acknowledgments}
This work was supported by the National Science Foundation under Grant No. CHE-2404788 and the Robert A. Welch Foundation (Grant No. E-1337).
\end{acknowledgments}







\end{document}